\documentclass[12pt]{article}
\usepackage{latexsym,amsfonts,amssymb}

\newtheorem{theorem}{Theorem}

\newtheorem{note}{Note}

\begin{document}

\title{Invariance of the quasilinear \\ equations of hiperbolic type  with respect\\
to the three-parametric Lie algebras}

\author{Olena Magda\thanks{e-mail: magda@imath.kiev.ua} \\ \small
Institute of Mathematics of the Academy of Sciences of Ukraine,\\
\small Tereshchenkivska Street 3, 01601 Kyiv-4, Ukraine}

\date{}

\maketitle

\begin{abstract}
We have  solved completely the problem of the description of
quasi-linear hy\-per\-bolic differential equations in two
independent variables that are invariant under three-parameter Lie
groups.
\end{abstract}

The problem of group classification of differential equations is one
of the central problems of  modern symmetry analysis of
differential equations[1]. One of the important classes of hyperbolic equations.
The problem of group classification of such equations has been discussed
by many authors (see for instance [2--9]).
 On this report we consider the problem of the
group classification of equations of form:
\begin{equation}\label{magda:eq1}
 u_{tt}=u_{xx}+F(t,x,u,u_x),
\end{equation}
where \ $u=u(t,x) and \ F$ is anarbitrary nonlinear differentiable function,  with\ $F_{u_{x},u_{x}}\not =0$ \
 is an arbitrary nonlinear smooth function,
which dependet variables \ $u$ \ or \ $u_x$. We use following notation
\ $u_x=\displaystyle{\frac{\partial u}{\partial x}}, \
u_{xx}=\displaystyle{\frac{\partial ^2u}{\partial x^2}}, \
F_{u_{x}}=\displaystyle{\frac{\partial F}{\partial u_x}}, \
u_t=\displaystyle{\frac{\partial u}{\partial t}}, \
u_{tt}=\displaystyle{\frac{\partial ^2u}{\partial t^2}}.$
For the group classification of equation (\ref{magda:eq1}) we use the approach proposed in [10].
Here we give three fundamental results (for details, the reader is refered to [11]).
\begin{theorem}
The infinitesimal operator of the symmetry group  of the equation (\ref{magda:eq1}) has following form:
\begin{equation}
X=(\lambda t+\lambda _1)\partial _t+(\lambda x+\lambda _2)\partial _x+
(h(x)u+r(t,x))\partial _u, \label{magda:eq2}
\end{equation}
where \ $\lambda , \lambda _1, \lambda  _2$are  arbitrary real constants
and $h(x), \ r(t,x)$ are  arbitrary functions  which satisfy the  condition
\begin{eqnarray}
r_{tt}-\displaystyle{\frac{d^2h}{dx^2}}u-r_{xx}+(h-2\lambda )F-(\lambda t+
\lambda _1)F_t-(\lambda x+\lambda _2)F_x- \nonumber \\
-(hu+r)F_u-2u_x\displaystyle
{\frac{dh}{dx}}-u_x(h-\lambda )F_{u_{x}}-\displaystyle{\frac{dh}{dx}}uF_{u_{x}}
-r_xF_{u_{x}}=0. \label{magda:eq3}
\end{eqnarray}
\end{theorem}
\begin{theorem}
The  equivalence group of the equation (\ref{magda:eq1}) is given
by  transformations  of the following  form:
\begin{equation}
\bar t=\gamma t+\gamma _1, \ \bar x=\epsilon \gamma x+\gamma _2, \ v=\rho (x)u+
\theta (t,x), \label{magda:eq4} \\
\gamma \not =0, \ \rho \not =0, \ \epsilon =\pm 1.\nonumber
\end{equation}
\end{theorem}
Using  theorem 2, one can  prove the
following result:
\begin{theorem}
In the class of operators (\ref{magda:eq2}) ,there are no realizations of
the algebras
 \ $so(3)$ \ and \ $sl(2,R)$.
\end{theorem}
From this theorem we obtain the following :

\begin{note}
 in the class of operators (\ref{magda:eq2}) there are no  realizations of any real
semi-simple Lie algebras;
\end{note}

\begin{note}
 there are no  equations (\ref{magda:eq1}) which has algebras of invariance, which are isomorphic
by real semi-simple algebras, or conclude those algebras as subalgebras.
\end{note}

The set of three-dimensional solvable Lie algebras consists of the following two decomposable
Lie algebras:
\[\!\!\!
\begin{array}{l}
A_{3.1}=A_1\oplus A_1\oplus A_1=3A_1;
\vspace{1mm}\\
A_{3.2}=A_{2.2}\oplus A_1,\quad [e_1,\ e_2]=e_2,
\vspace{1mm}\\
\end{array}
\]
and the following eight of non-decomposable Lie algebras:
\[\!\!\!
\begin{array}{l}
A_{3.3}: \quad [e_2,\ e_3]=e_1;
\vspace{1mm}\\
A_{3.4}: \quad [e_1,\ e_3]=e_1,\quad [e_2,\ e_3]=e_1+e_2;
\vspace{1mm}\\
A_{3.5}: \quad [e_1,\ e_3]=e_1,\quad [e_2,\ e_3]=e_2;
\vspace{1mm}\\
A_{3.6}: \quad [e_1,\ e_3]=e_1,\quad [e_2,\ e_3]=-e_2;
\vspace{1mm}\\
A_{3.7}: \quad [e_1,\ e_3]=e_1,\quad [e_2,\ e_3]=qe_2,\ (0<|q|<1);
\end{array}
\]
 We give the realizations of the algebras \ $A_{3.3}, \  A_{3.4},\ A_{3.5}, \
A_{3.6}, \ A_{3.7}, \ A_{3.8}, \ A_{3.9} $ \ and the
corresponding values of the functions \ $F$ \ in the equation
(\ref{magda:eq1})
\begin{eqnarray*}
 A_{3.3}^1 &=& \left\langle u\partial_{u},\ \partial_t+\beta\partial_x,\ {1\over \beta}xu\partial_{u},\right\rangle,
\beta >0:   \nonumber\\
&&F=-u^{-1}u_x^{2}+u \tilde G(\omega),\ \omega =x-\beta t; \nonumber\\
 A_{3.3}^2 &=& \left\langle u\partial_{u},\ \partial_x,\ m \partial _t+xu\partial_{u},\right\rangle,m\not =0:\nonumber\\
&&F=-u^{-1}u_x^{2}+u \tilde G(\omega),\ \omega =t-mu_xu^{-1}; \nonumber\\
 A_{3.3}^3 &=& \left\langle \partial_{u},\ \partial_x,\ m \partial _t+x\partial_{u},\right\rangle,m\not =0:\nonumber\\
&&F=\tilde G(\omega),\ \omega =mu_x-t; \nonumber\\
 A_{3.3}^4 &=& \left\langle \partial_{u},\ \partial_t,\  \partial _x+t\partial_{u},\right\rangle:\nonumber\\
&&F=\tilde G(u_x); \nonumber\\
A_{3.3}^5&=&\left\langle u\partial_{u},\ \partial_x,\  xu\partial_{u},\right\rangle:\nonumber\\
&&F=-u^{-1}u_x^{2}+u \tilde G(t); \nonumber\\
A^6_{3.3}& = &\langle t \partial_t +x \partial_x, u \partial_u, \ \ln |x|
u \partial_u \rangle: \nonumber\\
&&F=-u^{-1} u^2_x +\frac{1}{2}x^{-1} u_x +x^{-2} u\tilde G(\xi),\ \xi = tx^{-1}; \nonumber\\
 A_{3.3}^7 &=& \left\langle u\partial_{u},\ \partial_t+k \partial_x,\ m\partial _t+{1\over k }xu\partial_{u}\right\rangle,
k >0,\ m\in R:   \nonumber\\
&&F=-u^{-1}u_x^{2}+u \tilde G(\omega),\ \omega =x-k t+mku^{-1}u_x; \nonumber\\
A_{3.3}^8 &=&  \left\langle e^{mt} \partial_u,\ \partial _x,\ \partial _t+xe^{mt}\partial _u \right\rangle, m\not =0:\nonumber\\
&&F=m^2u+\tilde G(\omega), \ \omega ={e^{mt} \over m}-u_x;\nonumber\\
 A_{3.3}^9 &=& \left\langle \partial_{u},\ \partial_t,\  t\partial_{u},\right\rangle:\nonumber\\
&&F=\tilde G(x,u_x); \nonumber\\
 A_{3.3}^{10} &=& \left\langle u\partial_{u},\  \partial _t-\beta ^{-1}xu\partial_{u},\ \partial_t+\beta\partial_x \right\rangle,\beta >0:\nonumber\\
&&F=-u^{-1}u_x^{2}+u \tilde G(\omega),\ \omega =x-\beta t-\beta ^2u_xu^{-1}; \nonumber\\
 A_{3.3}^{11} &=& \left\langle u\partial_{u},\  \partial _t-xu\partial_{u},\ \partial_x \right\rangle:\nonumber\\
&&F=t^2u+2tu_x+u\tilde G(\omega),\ \omega =t+u_xu^{-1}; \nonumber\\
 A_{3.3}^{12} &=&  \left\langle e^{kt} \partial_u,\ \partial _t+ku\partial _u,\ \beta \partial _x+te^{kt}\partial _u \right\rangle, \beta>0,\ k>0:\nonumber\\
&&F=k^{2}u+{2kx \over \beta} e^{kt} +e^{kt} \tilde G(\omega), \ \omega =e^{-kt} u_x; \nonumber\\
A^{13}_{3.3} &=& \langle |t|^{\frac{1}{2}}\partial_u, -|t|^{\frac{1}{2}}\ln |t|\partial_u,\
t \partial_t +x \partial_x+\frac{1}{2}u\partial _u \rangle:  \nonumber\\
&&F=-\frac{u}{4} t^{-2}+u_x^{3} \tilde G(\omega ,v), \omega =tx^{-1},\ v=xu_x^{2};\nonumber\\
A^{14}_{3.3} &=& \langle \partial_u, -t\partial_u,\
 \partial_t +k \partial_x \rangle , k>0:  \nonumber\\
&&F=\tilde G(\omega ,u_x), \ \omega =x-kt; \nonumber\\A_{3.4}^1
&=& \left\langle \partial_{u},\ \partial_t+ k\partial_x,\
t\partial _t+x\partial _x +(u+ t)\partial_{u} \right\rangle,
\eta=x- kt,\ m=1,\ k>0:  \nonumber\\ &&F=\eta^{-1} \tilde
G(u_x);\nonumber\\
 A_{3.4}^ 2 &=& \left\langle \eta ^{m-1}\partial_{u},\ \partial_t+ k\partial_x,\ t\partial _t+x\partial _x
+(mu+ t\eta ^{m-1})\partial_{u} \right\rangle,\nonumber\\ &&\eta=x- kt,\ k>0,\ m \in R,\ m \not =1,2:  \nonumber\\
&&F=(k^2-1)(m-1)(m-2)\eta ^{-2}u-\frac{2k(1-m)}{2m-4}\eta ^{m-2}+\eta ^{2-m}\tilde G(\omega),\nonumber\\
&& \omega =((1-m)u+\eta u_x)\eta ^{3m-4};\nonumber\\
 A_{3.4}^3 &=& \left\langle \partial_{u},\ \partial_t,\ t\partial _t+x\partial _x
+(u+t)\partial_{u} \right\rangle:  \nonumber\\
&&F=x^{-1}\tilde G(u_x);\nonumber\\
 A_{3.4}^4 &=& \left\langle
e^{ktx^{-1}}\partial_{u},\ \partial _t+kx^{-1}u\partial _u,\
t\partial _t+x\partial _x +(u+ te^{ktx^{-1}})\partial_{u}
\right\rangle ,\ k\not =0:  \nonumber\\
&&F=u(k^{2}t^{2}x^{-4}-2ktx^{-3}+k^{2}x^{-2})+2ktu_xx^{-2}+e^{ktx^{-1}}(2k\ln
|x|x^{-1}+x^{-1}\tilde G(\omega)),\nonumber\\ &&\omega
=e^{-ktx^{-1}}(u_x+ktux^{-2});\nonumber\\
 A_{3.4}^5 &=& \left\langle kx^{-1}u\partial _u,\ -k\ln |x| x^{-1}u\partial _u,\
t\partial _t+x\partial _x  \right\rangle ,\ k>0:  \nonumber\\
&&F=-u^{-1}u_x^{2}+3xu_x+u\ln |u|x^{-2}+ux^{-2} \tilde G(\omega),\
\omega =tx^{-1};\nonumber\\
A_{3.4}^6 &=& \left\langle
kx^{\frac{1}{2}}u\partial _u,\ -k\ln |x|x^{\frac{1}{2}}u\partial
_u,\ 2(t\partial _t+x\partial _x  )\right\rangle ,\ k>0: \nonumber\\
&&F=-u^{-1}u_x^{2}+{1 \over 4}u\ln |u|x^{-2}+ux^{-2}
\tilde G(\omega),\ \omega =tx^{-1};\nonumber\\
A_{3.4}^7 & = & \langle -\partial_t +2\partial_x,\ e^{\frac{1}{2}x} u \partial_u,\
e^{\frac{1}{2}x}  \partial_u\rangle:  \nonumber\\
&&F=-u^{-1}u_x^{2}-u_x+{1 \over 4}u\ln |u|+u \tilde G(\eta),\ \eta=-2t-x;\nonumber\\
A_{3.4}^8 & = & \langle -2\partial_x, \
e^{\frac{1}{2}x} u \partial_u,\ e^{\frac{1}{2}x} xu
\partial_u\rangle:  \nonumber\\ &&F=-u^{-1}u_x^{2}-u_x+{1 \over 4}u\ln |u|+u \tilde G(t);\nonumber\\
 A_{3.4}^9 &=& \left\langle kx^{-1}u\partial _u,\ \partial _t-k\ln |x| x^{-1}u\partial _u,\
t\partial _t+x\partial _x  \right\rangle ,\ k>0:  \nonumber\\
&&F=k^2t^2ux^{-4}-3ktux^{-3}+2ktu_xx^{-2}+2ktux^{-3}\ln |u|\nonumber\\ &&-2ux^{-2}\ln |u|
+2u_xx^{-1}\ln |u|+x^{-2}u\ln ^2|u|+ux^{-2}\tilde G(\omega),\nonumber\\
&&\omega =xu_xu^{-1}+\ln |u|+ktx^{-1}\nonumber\\
A^{10}_{3.4} &=& \langle |t|^{\frac{1}{2}}\partial_u, -|t|^{\frac{1}{2}}\ln |t|\partial_u,\
t \partial_t +x \partial_x+\frac{3}{2}u\partial _u \rangle:  \nonumber\\
&&F=-\frac{u}{4} t^{-2}+u_x^{-1} \tilde G(\omega ,v), \omega =tx^{-1},\ v=x^{-1}u_x^{2};\nonumber\\
A^{11}_{3.4} &=& \langle \partial_u, -t\partial_u,\
 \partial_t +k \partial_x +u\partial _u\rangle , k>0:  \nonumber\\
&&F=u_x\tilde G(\omega ,v), \ \omega =x-kt, \ v=\ln |u_x|-t; \nonumber\\
 A_{3.5}^1 &=& \left\langle  \partial_{u},\ \partial_t+ k\partial_x,\ t\partial _t+x\partial _x
+u\partial_{u} \right\rangle, \eta=x- k t, \ k >0:  \nonumber\\
&&F=\zeta^{-1} \tilde G(u_x);\nonumber\\
 A_{3.5}^2 &=& \left\langle  \eta  \partial_{u},\ \partial_t+ k \partial_x,\  t\partial _t+x\partial _x
+2u\partial_{u} \right\rangle, \eta=x- k t,\ k >0:  \nonumber\\
&&F= \tilde G(\omega),\ \omega =(-u+u_x \eta)\eta ^{-2};\nonumber\\
 A_{3.5}^3 &=& \left\langle  \eta ^{m-1}  \partial_{u},\ \partial_t+ k\partial_x,\  t\partial _t+x\partial _x
+mu\partial_{u} \right\rangle, \eta=x- k t, \ k >0,\ m \in R,\ m\not =1,2:  \nonumber\\
&&F=(k^2-1)(m-1)(m-2)u \eta ^{-2}+\eta ^{m-2} \tilde G(\omega),\ \omega =((1-m)u+u_x \eta)\eta ^{-m};\nonumber\\
 A_{3.5}^4 &=& \left\langle \partial_x,\  \partial_{u},\ t\partial _t+x\partial _x
+u\partial_{u} \right\rangle:  \nonumber\\
&&F=t^{-1}\tilde G(u_x);\nonumber\\
 A_{3.5}^5 &=& \left\langle \partial_x,\  t\partial_{u},\ t\partial _t+x\partial _x
+2u\partial_{u} \right\rangle:  \nonumber\\
&&F=\tilde G(\omega),\ \omega =u_xt;\nonumber\\
 A_{3.5}^6 &=& \left\langle \partial_x,\  |t|^{m-1}\partial_{u},\ t\partial _t+x\partial _x
+mu\partial_{u} \right\rangle ,m\not =1,2,\ m\in R:  \nonumber\\
&&F=(2u-3mu-m^2u)t^{-2}+t^{m-2}\tilde G(\omega),\ \omega =u_xt^{m-1};\nonumber\\
  A_{3.5}^7 &=& \left\langle \partial_t,\  \partial_{x},\ t\partial _t+x\partial _x
 \right\rangle:  \nonumber\\
&&F=u_x^{2}\tilde G(u); \nonumber\\
 A_{3.5}^8 &=& \left\langle \partial_t,\  \partial_{x},\ t\partial _t+x\partial _x+mu\partial _u
 \right\rangle, \ m\not =0, 1,2:  \nonumber\\
&&F=|u|^{\frac{m-2}{m}}\tilde G(\omega), \omega =u_x^{-1}|u|^{\frac{m-1}{m}}; \nonumber\\
 A_{3.5}^9 &=& \left\langle \partial_t,\  \partial_{x},\ t\partial _t+x\partial _x+\partial _u
 \right\rangle, \ m\not =0:  \nonumber\\
&&F=e^{-2u}\tilde G(\omega),\ \omega =e^{u}u_x; \nonumber\\
 A_{3.5}^{10} &=& \left\langle \partial_t,\ x^{-1}u\partial _u,\  t\partial _t+x\partial _x \right\rangle ,k\not =0:  \nonumber\\
&&F=2u_xx^{-1} \ln |u|+u\ln^{2}|u|x^{-2}-2u\ln|u|x^{-2}+x^{-2}u\tilde G(\omega),\nonumber\\
&&\omega =u_xu^{-1}x+\ln|u|; \nonumber\\
A_{3.5}^{11} &=& \left\langle \partial _t+kx^{-1}u\partial _u,\  e^{ktx^{-1}}\partial_{u},\  t\partial _t+x\partial _x
+u\partial_{u} \right\rangle ,\ k\not =0:  \nonumber\\
&&F=uk(kt^{2}-2xt+kx^2)x^{-4}+2ktu_xx^{-2}+e^{ktx^{-1}}x^{-1}\tilde G(\omega),\nonumber\\
&&\omega =e^{-ktx^{-1}}(u_x+ktux^{-2});\nonumber\\
A_{3.5}^{12} &=& \left\langle \partial_t,\  \partial_{u},\ t\partial _t+x\partial _x
+u\partial_{u} \right\rangle:  \nonumber\\
&&F=x^{-1}\tilde G(u_x);\nonumber\\
 A_{3.6}^1 &=& \left\langle  \partial_t+k\partial_x,\ \partial_{u},\ t\partial _t+x\partial _x
-u\partial_{u} \right\rangle, \eta=x- k t,\ k >0:  \nonumber\\
&&F=\eta^{3} \tilde G(\omega), \ \omega =\eta ^{-2}u_x;\nonumber\\
 A_{3.6}^2 &=& \left\langle  \partial_t+k\partial_x,\ \eta \partial_{u},\ t\partial _t+x\partial _x
 \right\rangle, \eta=x- k t,\ k >0:  \nonumber\\
&&F=\eta^{2} \tilde G(\omega), \ \omega =-u+\eta u_x;\nonumber\\
 A_{3.6}^3 &=& \left\langle  \partial_t+k\partial_x,\ \eta ^{m+1}\partial_{u},\ t\partial _t+x\partial _x
+mu\partial_{u} \right\rangle, \eta=x- k t,\ k >0,\ m\in R,\ m\not =0,-1:  \nonumber\\
&&F=(k^2-1)m(m+1)u\eta ^{-2}+\eta^{2-m} \tilde G(\omega), \ \omega =(-(m+1)u+\eta u_x) \eta ^{3m};\nonumber\\
A_{3.6}^4 &=& \left\langle  \partial_{u},\ \partial_x,\ -t\partial _t-x\partial _x
+u\partial_{u} \right\rangle:  \ F=t^{-3}\tilde G(\omega), \ \omega =u_xt^2;\nonumber\\
A_{3.6}^5 &=& \left\langle   t\partial_{u},\ \partial_x,- t\partial _t-x\partial _x
+2u\partial_{u} \right\rangle:  \ F=t^{-4}\tilde G(\omega), \ \omega =u_xt^3;\nonumber\\
A_{3.6}^6 &=& \left\langle   t^{m-1}\partial_{u},\ \partial_x,- t\partial _t-x\partial _x
+mu\partial_{u} \right\rangle ,\ m\not =1,2,\ m\in R:  \nonumber\\
&&F=(2u-3mu+m^2u)t^{-2}+t^{-(m+2)}\tilde G(\omega), \ \omega =u_xt^{m+1};\nonumber\\
A_{3.6}^7 &=& \left\langle   \partial_x,\ t^{m+1}\partial_{u},\ t\partial _t+x\partial _x+mu\partial_{u}  \right\rangle ,\ m\not =0,-1,\ m\in R:  \nonumber\\
&&F=m(m+1)ut^{-2}+t^{m-2} \tilde G(\omega), \ \omega =u_xt^{1-m};\nonumber\\
A_{3.6}^8 &=& \left\langle   \partial_x,\ t\partial_{u},\ t\partial _t+x\partial _x \right\rangle:  \nonumber\\
&&F=t^{-2}\tilde G(\omega), \ \omega =u_xt;\nonumber\\
A_{3.6}^9 &=& \left\langle  \partial_x, \ \partial_{u},\ t\partial _t+x\partial _x
-u\partial_{u} \right\rangle:  \nonumber\\
&&F=t^{-3}\tilde G(\omega), \ \omega =u_xt^2;\nonumber\\
 A_{3.6}^{10} &=& \left\langle \partial_t,\  \partial_{u},\ t\partial _t+x\partial _x-u\partial_{u}
 \right\rangle:  \nonumber\\
&&F=x^{-2}\tilde G(\omega), \ \omega =u_xx^2; \nonumber\\
 A_{3.6}^{11} &=& \left\langle \partial_t,\ xu\partial _u,\  t\partial _t+x\partial _x \right\rangle ,k\not =0:  \nonumber\\
&&F=-2u_xx^{-1}\ln|u|+u\ln^{2}|u|x^{-2}+x^{-2}u\tilde G(\omega),\nonumber\\
&&\omega =u_xu^{-1}x-\ln|u|; \nonumber\\
 A_{3.6}^{12} &=& \left\langle \partial_t+mx^{-1}u\partial _u,\ xu\partial _u,\  t\partial _t+x\partial _x \right\rangle ,mk\not =0:  \nonumber\\
&&F=4m^2t^2x^{-4}u-2mtux^{-3}+4mtu_xx^{-2}-4mtux^{-3}\ln |u|-\nonumber\\ &&2u_xx^{-1}\ln |u|+u\ln ^2|u|x^{-2}+x^{-2}u \tilde G(\omega),  \nonumber\\
&&\omega =u_xu^{-1}x-\ln|u|+2mtx^{-1}; \nonumber\\
A_{3.6}^{13} &=& \left\langle \partial _t+kx^{-1}u\partial _u,\  e^{ktx^{-1}}\partial_{u},\  t\partial _t+x\partial _x
u\partial_{u} \right\rangle ,\ k\not =0:  \nonumber\\
&&F=uk(kt^{2}-2xt+kx^2)x^{-4}+2ktu_xx^{-2}+e^{ktx^{-1}}x^{-3}\tilde G(\omega),\nonumber\\
&&\omega =e^{-ktx^{-1}}(u_xx^2+ktu);\nonumber\\
A_{3.6}^{14} &=& \left\langle kx^{-1}u\partial _u,\ mxu\partial _u,\ t\partial _t+x\partial _x  \right\rangle ,\ mk\not =0:  \nonumber\\
&&F=-u^{-1}u_x^{2}+xu_x-u\ln |u|x^{-2}+ux^{-2} \tilde G(\omega),\ \omega =tx^{-1};\nonumber\\
A_{3.6}^{15} & = & \langle -\partial_t +\partial_x,\
e^{x} u \partial_u,\ e^{-x} u \partial_u\rangle:  \nonumber\\
&&F=-u^{-1}u_x^{2}-u\ln |u|+u \tilde G(\eta),\ \eta =-t-x;\nonumber\\
A_{3.6}^{16} & = &  \langle \partial_x, \ e^{x} u \partial_u,\ e^{-x} u \partial_u\rangle:  \nonumber\\
&&F=-u^{-1}u_x^{2}-\ln |u|+u \tilde G(t);\nonumber\\
A^{17}_{3.6} &=& \langle e^{(m-1)t}\partial_u, e^{(1-m)t}\partial_u,\
 \partial_t +k \partial_x +mu\partial _u\rangle , k\ge0,\ m\not =1:  \nonumber\\
&&F=u-2mu-m^2u+u_x \tilde G(\omega ,v), \ \omega =x-kt,\ v=\ln |u_x|-mt;\nonumber\\
\end{eqnarray*}
%\newpage
\begin{eqnarray*}
A_{3.7}^1 &=& \left\langle  \partial_t+ k\partial_x,\ \partial_{u},\ t\partial _t+x\partial _x
+qu\partial_{u} \right\rangle, \eta=x-k t,\ k >0,\  0<|q|<1:  \nonumber\\
&&F=\eta^{q-2} \tilde G(\omega), \ \omega =\eta ^{q-1}u_x^{-1};\nonumber\\
A_{3.7}^2 &=& \left\langle  \partial_t+ k\partial_x,\ \eta \partial_{u},\ t\partial _t+x\partial _x
+(q+1)u\partial_{u} \right\rangle, \eta=x-k t,\ k >0,\  0<|q|<1:  \nonumber\\
&&F=\eta^{q-1} \tilde G(\omega), \ \omega =\eta ^{-q} (u_x-u \eta ^{-1});\nonumber\\
A_{3.7}^ 3&=& \left\langle  \partial_t+ k\partial_x,\ \eta ^{m-q} \partial_{u},\ t\partial _t+x\partial _x
+mu\partial_{u} \right\rangle,\nonumber\\ && \eta=x-k t,\ k >0,\ m \in R,m \not =q,q+1,\ 0<|q|<1:  \nonumber\\
&&F=(k^2-1)(m-q)(m-q-1) \eta ^{-2} u+ \eta^{m-2} \tilde G(\omega), \ \omega =\eta ^{1-m}(u_x-(m-q)
u\eta ^{-1});\nonumber\\
A_{3.7}^4 &=& \left\langle \partial_x,\  \partial_{u},\ t\partial _t+x\partial _x
+qu\partial_{u} \right\rangle , \ 0<|q|<1:  \ F=t^{q-2}\tilde G(\omega), \ \omega =u_x^{-1}t^{q-1};\nonumber\\
A_{3.7}^5 &=& \left\langle \partial_x,\  t\partial_{u},\ t\partial _t+x\partial _x
+(q+1)u\partial_{u} \right\rangle , \ 0<|q|<1:  \nonumber\\
&&F=t^{q-1}\tilde G(\omega), \ \omega =u_x^{-1}t^{q};\nonumber\\
A_{3.7}^6 &=& \left\langle \partial_x,\  t^{m-q}\partial_{u},\ t\partial _t+x\partial _x
+mu\partial_{u} \right\rangle ,\ m\not =q,q+1,\ m\in R, \ 0<|q|<1:  \nonumber\\
&&F=(m-1-q)(m-q)t^{-2}u+t^{m-2}\tilde G(\omega), \ \omega =u_x^{-1}t^{m-1};\nonumber\\
 A_{3.7}^7 &=& \left\langle \partial_t,\  \partial_{u},\ t\partial _t+x\partial _x+qu\partial_{u}
 \right\rangle , \ 0<|q|<1:  \nonumber\\
&&F=x^{q-2}\tilde G(\omega), \ \omega =u_x^{-1}x^{q-1}; \nonumber\\
A_{3.7}^8 &=& \left\langle \partial_t,\ x^{-q}u\partial _u,\  t\partial _t+x\partial _x \right\rangle ,k\not =0, \ 0<|q|<1:  \nonumber\\
&&F=2qu_xx^{-1}\ln|u|+q^2u\ln^{2}|u|x^{-2}-(q+1)qx^{-2}u\ln |u|+x^{-2}u\tilde G(\omega),\nonumber\\
&&\omega =u_xu^{-1}x+q\ln|u|; \nonumber\\
 A_{3.7}^9 &=& \left\langle \partial_t+mx^{-1}u\partial _u,\ kx^{-q}u\partial _u,\  t\partial _t+x\partial _x \right\rangle ,mk\not =0, \ 0<|q|<1:  \nonumber\\
&&F=(m(q-1)t(m(q-1)tu+x((q+2)u-2u_xx))x^{-4}- \nonumber\\ && qx(2m(q-1)tu+x(u+qu-2u_xx))\ln |u|x^{-3}+q^2ux^{-2}\ln ^2|u|+ux^{-2}\tilde G(\omega),\nonumber\\
&&\omega =u_xux+q\ln |u|+m(1-q)tx^{-1} \nonumber\\
A_{3.7}^{10} &=& \left\langle \partial _t+kx^{-1}u\partial _u,\ e^{ktx^{-1}}\partial_{u},\  t\partial _t+x\partial _x
+qu\partial_{u} \right\rangle ,\ k\not =0, \ 0<|q|<1:  \nonumber\\
&&F=uk(kt^{2}x^{-4}-2tx+kx^{-2})+2ktu_xx^{-2}+e^{ktx^{-1}}+x^{q-2} \tilde G(\omega),\nonumber\\
&&\omega =e^{-ktx^{-1}}x^{1-q}(u_x+ktux^{-2});\nonumber\\
A_{3.7}^{11} &=& \left\langle kx^{-1}u\partial _u,\ mx^{-q}u\partial _u,\ t\partial _t+x\partial _x  \right\rangle ,\ mk\not =0, \ 0<|q|<1:  \nonumber\\
&&F=-u^{-1}u_x^{2}+(q+2)x^{-1}u_x+qu\ln |u|x^{-1}+ux^{-2} \tilde G(\omega),\ \omega =tx^{-1};\nonumber\\
A_{3.7}^{12} & = & \langle -\partial_t +2\partial_x,\
e^{\frac{1}{2}qx} u \partial_u,\ e^{\frac{1}{2}x} xu \partial_u\rangle, \ 0<|q|<1:  \nonumber\\
&&F=-u^{-1}u_x^{2}-\frac{q+1}{2}u_x+{q \over 4}u\ln |u|+u \tilde G(\eta),\ \eta =-2t-x;\nonumber\\
A_{3.7}^{13} & = & \langle -2\partial_x, \ e^{\frac{1}{2}x} u \partial_u,\ e^{\frac{1}{2}qx} xu \partial_u\rangle, \ 0<|q|<1: \nonumber\\
&&F=-u^{-1}u_x^{2}-\frac{q+1}{2}u_x+{q \over 4}u\ln |u|+u \tilde G(t);\nonumber\\
A^{14}_{3.7} &=& \langle e^{\frac{1}{2}(q+1)t}\partial_u, e^{\frac{1}{2}(1-q)t}\partial_u,\
 \partial_t +k \partial_x +\frac{1}{2}(1+q)u\partial _u\rangle , k\ge0,\ 0<|q|<1:  \nonumber\\
&&F=\frac{(q+1)^2}{4}u+u_x \tilde G(\omega ,v), \ \omega =x-kt,\ v=\ln |u_x|-\frac{q+1}{2}t;\nonumber\\
 A_{3.8}^1& =& \langle \cos (\ln |x|)   u \partial_u, \  \sin  (\ln |x|)   u \partial_u, \ t \partial_t +x \partial_x \rangle: \nonumber\\
&&F=-u^{-1}u_x^{2}+u_xx^{-1}+u\ln |u|x^{-2}+ux^{-2} \tilde G(\omega), \ \omega=tx^{-1}; \nonumber\\
 A_{3.8}^2& =& \langle \cos x  u \partial_u, \  \sin  x  u \partial_u, \  \partial_x \rangle: \nonumber\\
&&F=-u^{-1}u_x^{2}+u\ln |u|+u \tilde G(t); \nonumber\\
 A_{3.8}^3& =& \langle \cos x  u \partial_u, \  \sin  x  u \partial_u, \ \partial _t+ \partial_x \rangle: \nonumber\\
&&F=-u^{-1}u_x^{2}+u\ln |u|+u \tilde G(\eta),\ \eta =x-t; \nonumber\\
 A_{3.8}^4& =& \langle \cos t  \partial_u, \ - \sin  t \partial_u, \ \partial _t+k \partial_x \rangle ,\ k\ge 0: \nonumber\\
&&F=-u+ \tilde G(\eta ,u_x),\ \eta =x-kt; \nonumber\\
 A_{3.8}^5& =& \left \langle |t|^{\frac{1}{2}}\cos \left (\frac{\ln |t|}{2k}\right)   \partial_u, \  -|t|^{\frac{1}{2}}\sin  \left(\frac{\ln |t|}{2k}\right)   \partial_u,
\ k(2t \partial_t +2x \partial_x+u\partial _u)\right \rangle ,\ k\ge 0: \nonumber\\
&&F=-\frac{k^2+1}{4k^2}t^{-2}u+|t|^{-\frac{1}{2}} \tilde G(\omega ,v), \ \omega=tx^{-1},\ v=tu_x^2; \nonumber\\
 A_{3.9}^1& =& \langle x^{(q^2+1)}\cos ((q^2+1)\ln |x|)   u \partial_u, \   x^{(q^2+1)}\sin  ((q^2+1)\ln |x|)   u \partial_u,
 \ t \partial_t +x \partial_x, \rangle,\ q>0:  \nonumber\\
&&F=-u^{-1}u_x^{2}+(2q^2+1)u_xx^{-1}+2(q^2+1)^2u\ln |u|x^{-2}+ux^{-2} \tilde G(\omega), \ \omega=tx^{-1}; \nonumber\\
 A_{3.9}^2& =& \langle e^{\frac{1}{2}qx}\cos {x\over 2}  u \partial_u, \  e^{\frac{1}{2}qx}\sin { x\over 2}  u \partial_u,
\  -2\partial_x, \rangle,\ q>0: \nonumber\\
&&F=-u^{-1}u_x^{2}-qu_x+{(q^2+1)u\ln |u|\over 4}+u \tilde G(t); \nonumber\\
 A_{3.9}^3& =& \langle e^{\frac{1}{2}qx}\cos {x\over 2}  u \partial_u, \  e^{\frac{1}{2}qx}\sin { x\over 2}  u \partial_u,
\  -\partial _t+2\partial_x, \rangle,\ q>0: \nonumber\\
&&F=-u^{-1}u_x^{2}-qu_x+{(q^2+1)u\ln |u|\over 4}+u \tilde G(\eta), \ \eta =-2t-x; \nonumber\\
  A_{3.9}^4& =& \langle  \sin  t \partial_u, \ \cos t  \partial_u, \partial _t+k \partial_x+qu\partial _u \rangle ,\ k\ge 0,\ q>0: \nonumber\\
&&F=-u+ u_x\tilde G(\eta ,v),\ \eta =x-kt,\ v=e^{-qt}u_x; \nonumber\\
 A_{3.9}^5& =& \left \langle  \  |t|^{\frac{1}{2}}\sin  \left(\frac{\ln |t|}{2(k-q)}\right)   \partial_u, |t|^{\frac{1}{2}}\cos \left (\frac{\ln |t|}{2(k-q)}\right)   \partial_u,
\ 2(k-q)(t \partial_t +x \partial_x)+ku\partial _u\right \rangle ,\nonumber\\ &&\ k\in R,\ q>0,\ k\not =q : \nonumber\\
&&F=-\frac{(k-q)^2+1}{4(k-q)^2}t^{-2}u+|t|^{\frac{4q-3k}{2(k-q)}} \tilde G(\omega ,v), \ \omega=tx^{-1},\ v=|t|^{k-2q}|u_x|^{2(k-q)}; \nonumber\\
\end{eqnarray*}
\subsection*{Acknowledgements}
The author is grateful to   R.Z. Zhdanov and V.I. Lahno  for proposing
of the problem and for the help in the research.

\end{document}